\documentclass[12pt]{article}
\usepackage{amsmath,cite,epsfig,a4,times,euler,euscript}
\renewcommand{\baselinestretch}{1.1}
\newcommand{\myTitle}[1]{\begin{center}{\bf\Huge #1}\\[5ex]\end{center}}
\newcommand{\myAuthor}[1]{\begin{center}{\Large #1}\\[2ex]\end{center}}
\newcommand{\myAffiliation}[1]{\\[1ex]{\it\large #1}}
\newcommand{\myEmail}[1]{\\[1ex]{\tt\large #1}}
\newcommand{\myDate}{\begin{center}{\large\today}\\[5ex]\end{center}}
\newcommand{\myAbstract}[1]{\begin{center}\renewcommand{\baselinestretch}{1}{\bf Abstract}\\[2ex]\parbox{0.8\linewidth}{\small\hspace{15pt} #1}\end{center}\vspace{\baselineskip}}
\newcommand{\myReport}[1]{\hspace{\fill} #1}
\newcommand{\myPreprint}[1]{}
\newcommand{\myKeywords}[1]{}

\newcommand{\fudgea}{\\[0.3ex]}
\newcommand{\fudgeb}{\\}%[-0.6ex]}
\newcommand{\Appendix}[1]{Appendix~\ref{#1}}   
\newcommand{\Section}[1]{Section~\ref{#1}}
\newcommand{\Table}[1]{Table~\ref{#1}}

\newcommand{\ie}{{\it i.e.}}
\newcommand{\graph}[3]{\raisebox{-#3pt}{\epsfig{file=#1.eps,width=#2pt}}}
\newcommand{\Kaleu}{\mbox{{\sc Kaleu}}}

\newcommand{\Alpgen}{{\sc Alpgen}}
\newcommand{\Alpcite}{\cite{Mangano:2002ea}}
\newcommand{\qmom}{q}
\newcommand{\pmom}{p}
\newcommand{\Qmom}{Q}
\newcommand{\zvar}{z}
\newcommand{\m}{-}
\newcommand{\p}{+}
\newcommand{\Nacc}{N_{\mathrm{eval}}}
\newcommand{\GeV}{\mathrm{GeV}}
\newcommand{\stens}{\sqrt{10}\sigma}
\newcommand{\ones}{1\sigma}
\newcommand{\AmHel}{\cite{Gleisberg:2003bi}}
\newcommand{\Carlo}{\cite{Kolodziej:2009cx}}
\newcommand{\HelPhe}{{\sc H/Phegas}}
\newcommand{\HelKal}{{\sc H/Kaleu}}

\begin{document}

\myReport{IFJPAN-IV-2010-3}
\myPreprint{IFJPAN-\\arXiv:}

\myTitle{%
\Kaleu: a general-purpose\fudgea
parton-level phase space generator%
\footnotetext{%
This work was partially supported by RTN European Programme, MRTN-CT-2006-035505 (HEPTOOLS, Tools and Precision Calculations for Physics Discoveries at  Colliders)
}
}

\myAuthor{%
A.\ van Hameren%
\myAffiliation{%
The H.\ Niewodnicza\'nski Institute of Nuclear Physics \fudgeb
Polisch Academy of Sciences\\
Radzikowskiego 152, 31-342 Cracow, Poland%
\myEmail{hameren@ifj.edu.pl}
}
}

\myDate

\myAbstract{%
\Kaleu\ is an independent, true phase space generator. After
providing it with some information about the field theory and the
particular multi-particle scattering process under consideration, it
returns importance sampled random phase space points. Providing it
also with the total weight of each generated phase space point, it
further adapts to the integration problem on the fly. It is written
in {\sc Fortran}, such that it can independently deal with several
scattering processes in parallel.
}

\myKeywords{}

%\begin{document}
%

%%%%%%%%%%%%%%%%%%%%%%%%%%%%%%%%%%%%%%%%%%%%%%%%%%%%%%%%%%%%%%%%%%%%%%%%%%%%%%%%
\section{Introduction\label{Sec:intro}}
%%%%%%%%%%%%%%%%%%%%%%%%%%%%%%%%%%%%%%%%%%%%%%%%%%%%%%%%%%%%%%%%%%%%%%%%%%%%%%%%
Efficient phase space generation is an important issue in the study of multi-particle processes at collider experiments.
Parton-level defined observables are eventually formulated as phase space integrals, and the evaluation of these is the most time-consuming part of a calculation.
The complicated structure of both the integrand and massive multi-particle phase space enforces the use of Monte-Carlo methods, setting the total evaluation time to the sum of the generation of phase space points and the evaluation of the integrand at the phase space points.
While it is important to keep the computing time of each integrand evaluation as low as possible, it is also important to keep the number of evaluations to reach an acceptable accuracy  as low as possible.
The latter is commonly referred to as {\em efficient phase space generation}, and eventually is a trade off between the number of evaluations and the average computing time for a single generation of a phase space point.

Since efficient phase space generation should already be settled at leading order (LO) calculations, automatic systems for LO calculations developed over the last decade deal with this issue
\cite{%
Boos:1994xb,
Yuasa:1999rg,
Papadopoulos:2000tt,
Krauss:2001iv,
Dittmaier:2002ap,
Mangano:2002ea,
Moretti:2001zz,
Kilian:2001qz,
Maltoni:2002qb,
Kersevan:2004yg
}.
Various techniques are used in various combinations.
Efficient phase space generation implies knowledge about the integrand, which is utilized both before the integral is performed, through phase space mappings, and during the integration process, using the actual value of the integrand at the phase space points.
Methods using the latter are referred to as {\em adaptive\/} methods, and the ones used within particle physics are based on adaptive grids dividing phase space into subspaces
\cite{%
Lepage:1977sw,
Kawabata:1995th,
Jadach:2002kn,
Hahn:2005pf,
vanHameren:2007pt
}, 
concern the optimization of the parameters in mixture distributions (the so-called {\em adaptive multi-channel method})
\cite{%
Kleiss:1994qy
},
or combine both
\cite{%
Ohl:1998jn
}.
Most important is, however, to use as much as possible information about the integrand before performing the integral, be it just when deciding which underlying phase space mapping to use in adaptive grid methods.

Eventually, the aim of the game is to have the probability density with which phase space points are generated look as much as possible as the integrand and to let it have the same peak structure.
The fact that, for parton-level multi-particle calculations, this structure is essentially determined by the sum of the squares of the tree-level Feynman graphs contributing to the scattering amplitude is of great help.
Since each Feynman graph encodes the phase space mapping needed for the efficient integration of its own square, the multi-channel method, in which each graph represents a channel, can be applied for efficient performance of the full integration.
The problem with this approach is that the number of Feynman graphs grows factorially with the number of final-state particles.
This problem was solved for the calculation of the tree-level amplitudes themselves through the use of recursive techniques
\cite{%
Berends:1987me,
Caravaglios:1995cd,
Draggiotis:1998gr,
Caravaglios:1998yr,
Kanaki:2000ey,
Moretti:2001zz
}.
Regarding the phase space generation, the straightforward solution within the multi-channel method is to keep only a restricted number of channels with the highest channel weight.
However, in order to determine these, the integration process still has to be started including all channels.

The problem only exists in the calculation of the weight coming with each phase space point.
Only there the sum over all graphs occurs, not in the actual generation of the phase space points, for each of which only one graph encoding a phase space mapping is chosen.
The reason why the weight cannot be calculated with the same recursive approach like the amplitude is that, within the multi-channel method, each individual graph has its own unique weight.
And indeed, the problem is solved by departing from this approach, as was pointed out in
\cite{%
Gleisberg:2008fv
},
and instead giving individual weights to the vertices.
Instead of choosing a whole graph encoding a mapping at once for each generation of a phase space point, a mapping is composed by choosing respective branchings.
Each branching corresponds to a vertex, and has its unique weight among the list of possible branchings at a given stage in the construction of the phase space mapping.
The weight associated with the generated phase space points can then be calculated following recursion relations analogous to those for the amplitude calculation.

As said, the foregoing works well assuming the peak structure of the integrand is determined by sum of the squares of the tree-level Feynman graphs contributing to the scattering amplitude, which however is not a gauge-invariant quantity.
For processes involving many gluons, other phase space mappings have been designed
\cite{%
Draggiotis:2000gm,
vanHameren:2000aj,
vanHameren:2002tc
}.

Although many implementations of the various methods for efficient phase space generation exist, most of them are heavily integrated in automatic systems for full multi-particle calculations, making it difficult to use them separately.
This write-up presents a program with which Kinematics Are Lucidly and Efficiently Utilized%
\footnote{ \Kaleu\ outranks {\sc Sarge}.}.
It makes a number of routines available which can be used for phase space generation in any given Monte Carlo program.
As part of the initialization, the user has to provide masses and widths of all possible real and virtual particles in the multi-particle process under consideration, and a list of interaction vertices.
\Kaleu\ uses this information to efficiently map phase space following the method of the Recursive Phase space Generator described in
\cite{%
Gleisberg:2008fv
}.
The generator is optimized on the fly during the integration process, with the option to also optimize the generation of invariants and polar angles further with {\sc Parni}
\cite{%
vanHameren:2007pt
}.
It is a {\sc Fortran} program, written such that  several instances of the program can operate in parallel, completely independently of each other.
It can be obtained from
\begin{center}
{\tt http://annapurna.ifj.edu.pl/\~{}hameren}
\end{center}

%%%%%%%%%%%%%%%%%%%%%%%%%%%%%%%%%%%%%%%%%%%%%%%%%%%%%%%%%%%%%%%%%%%%%%%%%%%%%%%%
\section{The algorithms\label{Sec:Alg}}
%%%%%%%%%%%%%%%%%%%%%%%%%%%%%%%%%%%%%%%%%%%%%%%%%%%%%%%%%%%%%%%%%%%%%%%%%%%%%%%%
As mentioned before, \Kaleu\ uses the algorithm of the Recursive Phase space Generator from
\cite{%
Gleisberg:2008fv
}.
Like most algorithms, it uses the fact that $n$-body phase space can be decomposed in a sequence of two-body phase spaces, referred to as {\em splittings} in the following.
%

%%%%%%%%%%%%%%%%%%%%%%%%%%%%%%%%%%%%%%%%%%%%%%%%%%%%%%%%%%%%%%%%%%%%%%%%%%%%%%%%
\subsection{Two-body phase space generation}
%%%%%%%%%%%%%%%%%%%%%%%%%%%%%%%%%%%%%%%%%%%%%%%%%%%%%%%%%%%%%%%%%%%%%%%%%%%%%%%%
In each splitting, two new four-momenta $\pmom_1,\pmom_2$ are generated that sum up to an, possibly previously generated, existing four-momentum $\Qmom=\pmom_1+\pmom_2$.
Depending on whether 0, 1 or 2 of these momenta are on-shell, 2, 3 or 4 random variables have to be generated from which $\pmom_1,\pmom_2$ are constructed.
Most common is to generate the invariants $\pmom_1^2,\pmom_2^2$, the cosine $\zvar$ of the polar angle of $\vec{\pmom}_1$ in the center-of-mass frame (CMF) of $\Qmom$, and the the azimuthal angel $\phi$ in the same frame.
If any of the momenta is on-shell, and less than 4 random variables have to be generated, we say ``the invariant is generated with a Dirac delta-distribution''.

The azimuthal angle is usually not correlated with any existing kinematical variable, but the $\zvar$-variable sometimes is.
It may be the cosine of the polar angle w.r.t.\ to another, already generated, momentum in the CMF of $\Qmom$.
The $\zvar$-variable is then linearly related to the scalar product of the two momenta.
This happens, for example, in the so called {\em open antenna\/} generation in
\cite{%
vanHameren:2002tc
}
in order to arrive at antenna densities for the efficient integration of multi-gluon amplitudes.
It also happens in the the so-called $t$-type generation
\cite{%
Byckling:1973
}.
In this case, the $\zvar$-variable is taken to be the cosine of the polar angle w.r.t.\ to one of the initial-state momenta $\qmom_1$ in the CMF of $\Qmom$, and is linearly related to $t=(\pmom_1-\qmom_1)^2=(\Qmom-\qmom_1-\pmom_2)^2$.
If this generation is formulated as $(\qmom_1,\Qmom')\to(\pmom_1,\pmom_2)$ with $\Qmom'=\Qmom-\qmom_1$, these four momenta can be interpreted as the external legs of a $t$-channel Feynman graph.
This is useful for bookkeeping purposes, as it will also be for \Kaleu. 
Both this $t$-type generation and the ``$s$-type'' generation with uncorrelated $z$ variable are applied in \Kaleu.

The mentioned different choices of random variables are applied in order to have control over the densities following which they are distributed.
They should match the behavior of the the squared matrix element one is trying to integrate.
\Kaleu\ uses the densities as mentioned for example in
\cite{%
Papadopoulos:2000tt
}.

%%%%%%%%%%%%%%%%%%%%%%%%%%%%%%%%%%%%%%%%%%%%%%%%%%%%%%%%%%%%%%%%%%%%%%%%%%%%%%%%
\subsection{Recursive phase space decomposition}
%%%%%%%%%%%%%%%%%%%%%%%%%%%%%%%%%%%%%%%%%%%%%%%%%%%%%%%%%%%%%%%%%%%%%%%%%%%%%%%%
We will illustrate how sequences of splittings are composed with the help of an explicit example, namely the process $u\Bar{u}\to{d}\Bar{d}Z$.
For this process, \Kaleu\ will create the list of splittings/mergings, or just {\em vertices\/}, in \Table{Table1}. 
%
%\begin{center}
\begin{table}
{\small
\begin{tabular}{|rccccc|}
\hline
%    &          &                       &         &         &          \\
   1:&  $g( 6)$ & $\longleftrightarrow$ & $d( 2)$ & $d( 4)$ &          \\
   2:&  $A( 6)$ & $\longleftrightarrow$ & $d( 2)$ & $d( 4)$ &          \\
   3:&  $Z( 6)$ & $\longleftrightarrow$ & $d( 2)$ & $d( 4)$ &          \\
   4:&  $d(10)$ & $\longleftrightarrow$ & $d( 2)$ & $Z( 8)$ &          \\
   5:&  $d(12)$ & $\longleftrightarrow$ & $d( 4)$ & $Z( 8)$ &          \\
   6:&  $u( 7)$ & $\longleftrightarrow$ & $W( 5)$ & $d( 2)$ & $[d( 4)]$  \\
   7:&  $u( 7)$ & $\longleftrightarrow$ & $W( 3)$ & $d( 4)$ & $[d( 2)]$  \\
   8:&  $W(11)$ & $\longleftrightarrow$ & $u( 9)$ & $d( 2)$ & $[Z( 8)]$  \\
   9:&  $W(11)$ & $\longleftrightarrow$ & $W( 3)$ & $Z( 8)$ & $[d( 2)]$  \\
  10:&  $W(13)$ & $\longleftrightarrow$ & $u( 9)$ & $d( 4)$ & $[Z( 8)]$  \\
  11:&  $W(13)$ & $\longleftrightarrow$ & $W( 5)$ & $Z( 8)$ & $[d( 4)]$  \\
  12:&  $g(14)$ & $\longleftrightarrow$ & $d( 2)$ & $d(12)$ &          \\
  13:&  $A(14)$ & $\longleftrightarrow$ & $d( 2)$ & $d(12)$ &          \\
  14:&  $Z(14)$ & $\longleftrightarrow$ & $d( 2)$ & $d(12)$ &          \\
  15:&  $g(14)$ & $\longleftrightarrow$ & $d( 4)$ & $d(10)$ &          \\
  16:&  $A(14)$ & $\longleftrightarrow$ & $d( 4)$ & $d(10)$ &          \\
  17:&  $Z(14)$ & $\longleftrightarrow$ & $d( 4)$ & $d(10)$ &          \\
\end{tabular}
\hspace{\fill}
\begin{tabular}{|rccccc|}
%    &          &                       &         &         &          \\
  18:&  $u(15)$ & $\longleftrightarrow$ & $u( 1)$ & $g(14)$ &          \\
  19:&  $u(15)$ & $\longleftrightarrow$ & $u( 1)$ & $A(14)$ &          \\
  20:&  $u(15)$ & $\longleftrightarrow$ & $u( 1)$ & $Z(14)$ &          \\
  21:&  $u(15)$ & $\longleftrightarrow$ & $W(13)$ & $d( 2)$ &          \\
  22:&  $u(15)$ & $\longleftrightarrow$ & $W(13)$ & $d( 2)$ & $[d(12)]$  \\
  23:&  $u(15)$ & $\longleftrightarrow$ & $W(11)$ & $d( 4)$ &          \\
  24:&  $u(15)$ & $\longleftrightarrow$ & $W(11)$ & $d( 4)$ & $[d(10)]$  \\
  25:&  $u(15)$ & $\longleftrightarrow$ & $u( 7)$ & $Z( 8)$ &          \\
  26:&  $u(15)$ & $\longleftrightarrow$ & $u( 7)$ & $Z( 8)$ & $[Z( 6)]$  \\
  27:&  $u(15)$ & $\longleftrightarrow$ & $u( 7)$ & $Z( 8)$ & $[A( 6)]$  \\
  28:&  $u(15)$ & $\longleftrightarrow$ & $u( 7)$ & $Z( 8)$ & $[g( 6)]$  \\
  29:&  $u(15)$ & $\longleftrightarrow$ & $W( 3)$ & $d(12)$ & $[d( 2)]$  \\
  20:&  $u(15)$ & $\longleftrightarrow$ & $W( 5)$ & $d(10)$ & $[d( 4)]$  \\
  31:&  $u(15)$ & $\longleftrightarrow$ & $u( 9)$ & $g( 6)$ & $[Z( 8)]$  \\
  32:&  $u(15)$ & $\longleftrightarrow$ & $u( 9)$ & $A( 6)$ & $[Z( 8)]$  \\
  33:&  $u(15)$ & $\longleftrightarrow$ & $u( 9)$ & $Z( 6)$ & $[Z( 8)]$  \\
%    &          &                       &         &         &          \\
\hline
\end{tabular}
}
\caption{List of vertices for the process $u\Bar{u}\to{d}\Bar{d}Z$}
\label{Table1}
\end{table}
%\end{center}
%
The list contains almost all possible three-point vertices occurring in Feynman graphs for this process, \ie\ all three-point vertices to be calculated in a recursive calculation of the amplitude.
A few are missing, as will be explained below, and a few appear several times with differences in the last column, which will also be explained.
The list only contains three-point vertices, the meaning of the particles between square brackets is explained below.
There is no distinction between particles that only differ in charge since it does not matter for the kinematics.
The number between parentheses encodes the momentum in the binary representation.
The external particles have momenta $u(1)u(16)\to{d}(2){d}(4)Z(8)$, and for example the gluon/photon/$Z$-boson attached to the ${d}(2){d}(4)$ pair has momentum $2+4=6$.
For a process with $n$ particles in the final state, momentum conservation dictates that initial-state momentum~$2^{n+1}$ is equal to minus momentum~$2^{n+1}-1=1+2+4+\ldots+2^{n}$.

To generate a phase space point, the list should be read upside down, and each vertex represents a splitting.
The starting point is the sum of all final-state momenta $2+4+8$, plus one initial-state momentum~$1$.
The latter is necessary to include $t$-type kinematics.
Thus a vertex is chosen among number~$18$ to~$33$.
Each of these carries its own multi-channel weight, which is updated during the Monte Carlo process.
Suppose vertex~$29$ is chosen, having $W( 3)$ , $d(12)$ on the r.h.s.
Then the same is repeated for all vertices with~$W( 3)$ on the l.h.s.\ and with~$d(12)$ on the l.h.s..
For the latter, this involves only vertex~$5$.
For the former, no such vertex exists in the list, because it would represent the trivial operation of subtracting the initial-state momentum~$1$ from~$3$ to obtain the final-state momentum~$2$.
Thus the constructed Feynman graph representing the phase space mapping is given by
\begin{equation}
\graph{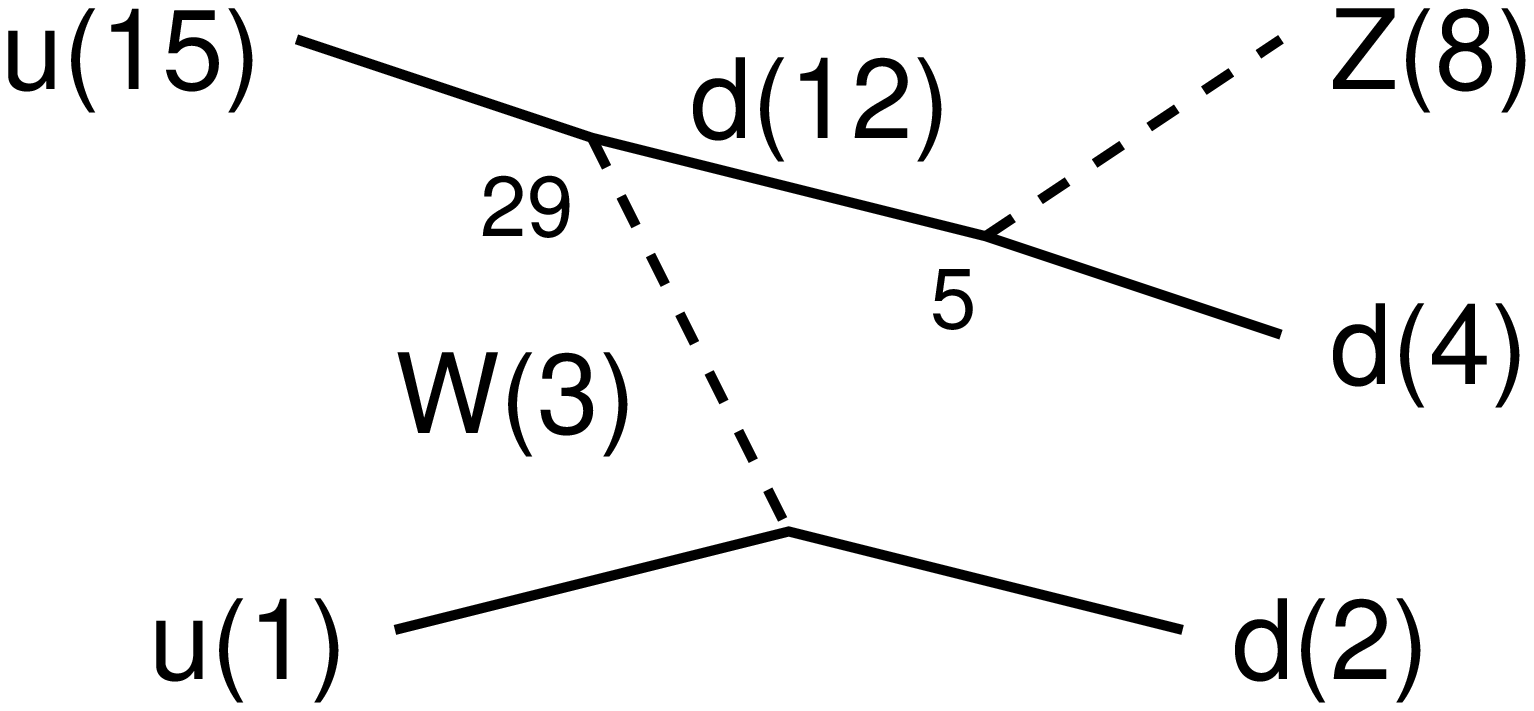}{160}{40}
\end{equation}
Vertex~$29$ represents a $t$-type splitting, as all vertices do with an odd momentum on the l.h.s., because they contain momentum~$1$.
The first step in a $t$-type splitting is the generation of the positive invariants, in the example above the squares of momenta~$12$ and~$2$.
The latter is actually external and fixed to the squared mass of the $d$-quark, so it is ``generated following a delta-distribution''.
One of these invariants, in this case the square of momentum~$2$, is beyond the encoding of the three-point vertex, and has to be listed separately.
This is the purpose of the particles between square brackets in the list of vertices.
Splittings of the $s$-type, like vertex~$12$, do not have entries in this last column, as well as $t$-type vertices which give the possibility to be followed by another $t$-type splitting, like vertex~$25$.
\begin{equation}
\graph{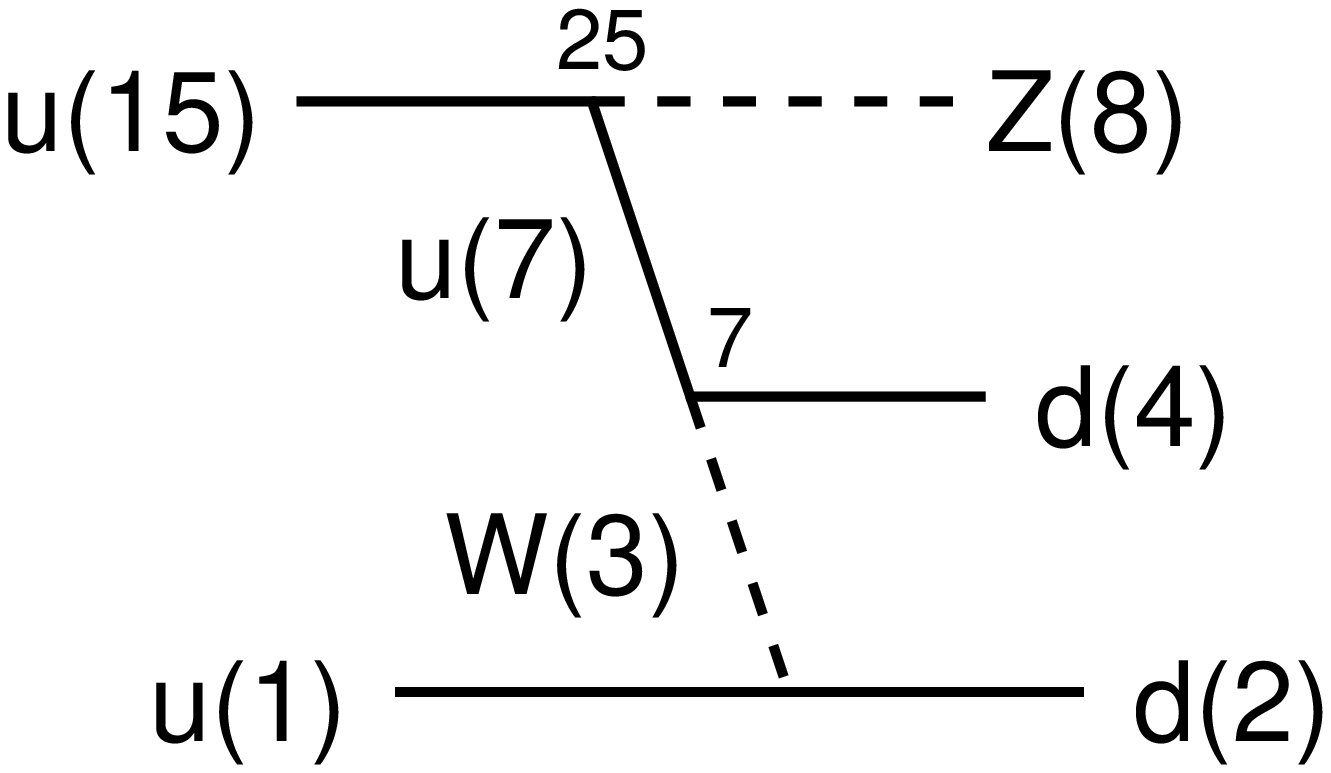}{145}{40}
\end{equation}
As mentioned before, vertices representing a splitting involving momentum~$1$ explicitly are omitted.
The only exception are vertices with the sum of all final-state momenta on the other leg, \ie\ vertices~$18$, $19$ and~$20$, since they are the starting point for pure $s$-type Feynman graphs like
\begin{equation}
\graph{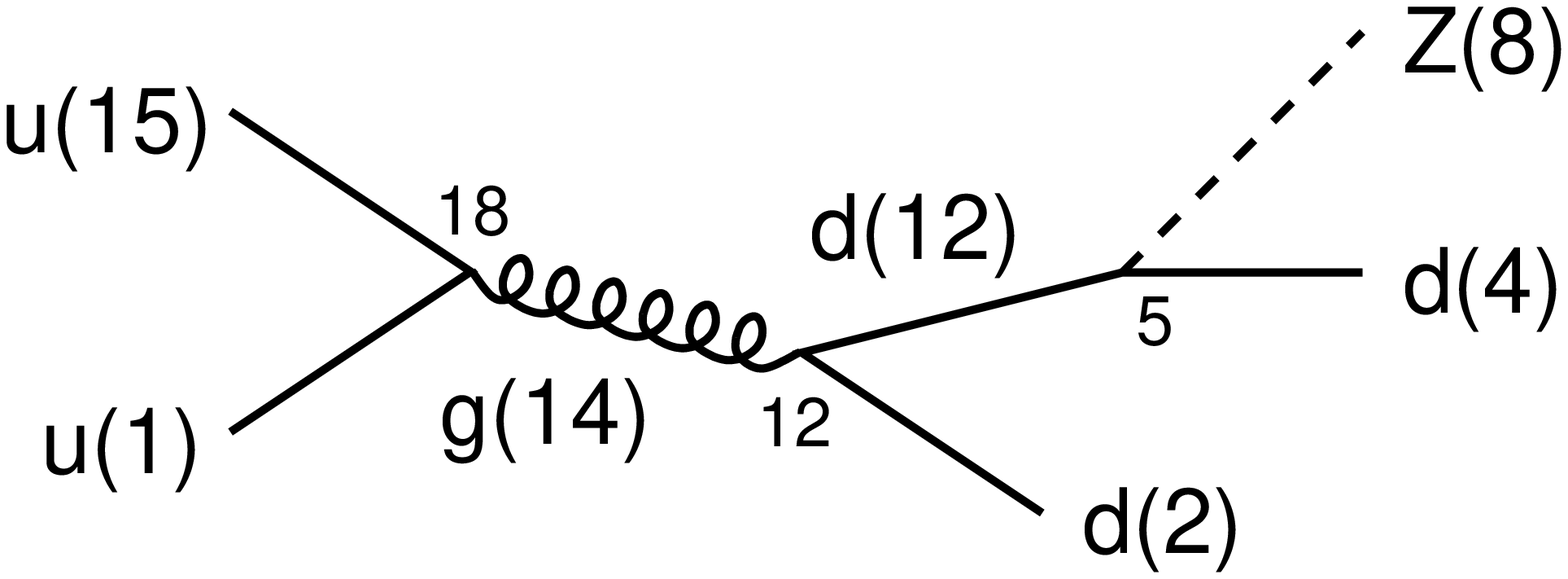}{210}{40}
\end{equation}

For the calculation of the weight associated with a phase space pointed generated as above, the list of vertices has to be read in the order as it stands.
Each ``$\longleftrightarrow$'' should now be interpreted as~``$=\!+$'' \ie\ each vertex represents a contribution to the ``off-shell current'' labeled by particle-type and momentum on the l.h.s..
These off-shell currents are probability densities, and the final density labeled by $u(15)$, obtained after executing the whole list of vertices, is the reciprocal of the weight.
Each vertex includes the density associated with the variables generated for the splitting, and the multi-channel weight for the choice of the vertex among the possibilities contributing to the same off-shell current.
This way, the contribution of all Feynman graphs is included, underlining the recursive character of the algorithm.

%%%%%%%%%%%%%%%%%%%%%%%%%%%%%%%%%%%%%%%%%%%%%%%%%%%%%%%%%%%%%%%%%%%%%%%%%%%%%%%%
\section{Usage\label{Sec:Use}}
%%%%%%%%%%%%%%%%%%%%%%%%%%%%%%%%%%%%%%%%%%%%%%%%%%%%%%%%%%%%%%%%%%%%%%%%%%%%%%%%
\Kaleu\ has been designed such that the user can conveniently write his own user-defined routines suitable for a given Monte Carlo program.
Some examples are included in the program.
In particular, an interface to replace {\sc Phegas} with \Kaleu\ in the {\sc Helac/Phegas} system
\cite{%
Kanaki:2000ey,
Papadopoulos:2000tt,
Cafarella:2007pc
}
is included.

The first thing the user has to do is provide \Kaleu\ with all particles and possible vertices.
For the example in \Section{Sec:Alg} the following routine would be sufficient:

{\scriptsize
\begin{verbatim}
  subroutine my_model( model ,vertx )
  use avh_kaleu_model
  type(model_type) ,intent(out) :: model
  type(vertx_type) ,intent(out) :: vertx
  integer :: gluon,photon,wboson,zboson,uquark,dquark
!
  parameter(  gluon=1 )
  parameter( photon=2 )
  parameter( wboson=3 )
  parameter( zboson=4 )
  parameter( uquark=5 )
  parameter( dquark=6 )
!
  call addparticle( model , gluon ,'g ' ,0d0      ,0d0     )
  call addparticle( model ,photon ,'A ' ,0d0      ,0d0     )
  call addparticle( model ,wboson ,'W ' ,80.419d0 ,2.048d0 )
  call addparticle( model ,zboson ,'Z ' ,91.188d0 ,2.446d0 )
  call addparticle( model ,uquark ,'u ' ,0d0      ,0d0     )
  call addparticle( model ,dquark ,'d ' ,0d0      ,0d0     )
!
  call addvertex( vertx ,uquark,uquark,gluon )
  call addvertex( vertx ,dquark,dquark,gluon )
  call addvertex( vertx ,uquark,uquark,photon )
  call addvertex( vertx ,dquark,dquark,photon )
  call addvertex( vertx ,uquark,uquark,zboson )
  call addvertex( vertx ,dquark,dquark,zboson )
  call addvertex( vertx ,uquark,dquark,wboson )
  call addvertex( vertx ,wboson,wboson,zboson )
  call addvertex( vertx ,wboson,wboson,photon )
  call addvertex( vertx , gluon, gluon,gluon )
!
  end subroutine
\end{verbatim}
}
\noindent The output \verb|model| and \verb|vertx| are of public derived type with private components, so the user can only declare them and pass them to routines from \Kaleu, in this case \verb|addparticle| and \verb|addvertex|.
The input of these routines should be clear, it should only be mentioned that the order of the particles in \verb|addvertex| does not matter.
In practise, the user would of course extend the model to, for example, the full standard model, which was refrained from for brevity here.
Some prepared model routines are included in the program.

A routine as above would typically be called in a user-defined initialization routine to be called in the Monte Carlo program.
A complete set of such user-defined routines could look as follows:

{\scriptsize
\begin{verbatim}
module my_kaleu
  use avh_kaleu
  use avh_kaleu_model
  use avh_kaleu_kinem
  type(model_type) ,save :: mdl ! Masses and widths
  type(kinem_type) ,save :: kin ! Kinematics
  type(kaleu_type) ,save :: obj ! Instance of the phase space generator
end module
      
subroutine my_init( process ,nfinst ,ecm ,nbatch,nstep,thrs )
  use my_kaleu
  integer ,intent(in)         :: process(-2:17),nfinst,nbatch,nstep
  real(kind(1d0)) ,intent(in) :: ecm,thrs
  type(vertx_type) :: vtx
  call my_model( mdl,vtx )
  call kaleu_put_process( mdl,vtx,kin,obj ,process ,nfinst ,ecm )
  call kaleu_init_adapt( mdl,obj ,nbatch,nstep,thrs )
end subroutine
  
subroutine my_gnrt( discard ,pkaleu )
  use my_kaleu
  logical         ,intent(out) :: discard
  real(kind(1d0)) ,intent(out) :: pkaleu(0:3,-2:17)  
  call kaleu_gnrt( mdl,kin,obj ,discard ,pkaleu )
end subroutine

subroutine my_wght( weight )
  use my_kaleu
  real(kind(1d0)) ,intent(out) :: weight
  call kaleu_wght( mdl,kin,obj ,weight )
end subroutine

subroutine my_collect( weight )
  use my_kaleu
  real(kind(1d0)) ,intent(in) :: weight
  call kaleu_collect( obj ,weight )
end subroutine
\end{verbatim}
}

The module \verb|my_kaleu| contains essentially what in a traditional {\sc Fortran} program would be the common blocks.
The items are of public derived type with private components, so the user can only declare them and pass them to routines from \Kaleu.
By declaring them as arrays, several instances of the program can be created.
The user-defined routines above are wrappers of \Kaleu-routines which need the items in \verb|my_module| as arguments.

The initialisation routine \verb|my_init| takes the process as input, and the particles should have the same integer labelling as in the user-defined \verb|my_model|.
The entries $-2$ and $-1$ of \verb|process| are the initial-state particles, the (strictly) positive entries contain the final-state particles.
The integer \verb|nfinst| is the number of final-state particles.
The input \verb|ecm| is the center-of-mass energy.
The numbers \verb|nbatch|, \verb|nstep| and \verb|thrs| are related to the multi-channel optimization.
This happens in \verb|nstep| steps of \verb|nbatch| non-zero weight events.
After these \verb|nstep| steps, each channel with a weight smaller than \verb|thrs| times the average channel weight at the off-shell current it belongs to is discarded.

\Kaleu\ determines limits on two-particle invariants based on the masses of final-state particles, but these may be equal zero and in a practical calculation there will be phase space cuts present.
The user may translate these into two-particle invariant mass cuts, and feed them to \Kaleu\ for more efficient phase space generation with
\begin{verbatim}
  call kinem_updt_smin( kin ,smin )
\end{verbatim}
The entries \verb|smin(1:nfinst,1:nfinst)| of the array \verb|smin(-2:17,-2:17)| should be minima to final-final state two-particle invariants, and the entries \verb|smin(-2:-1,1:nfinst)| should be negative, and should be maxima to final-initial state two-particle invariants.
The array is considered to be symmetric.

All arguments of the generation routine \verb|my_gnrt| are output.
If the logical \verb|discard| is true, the phase space point should get weight zero, and the array \verb|pkaleu| must not be used.
If \verb|discard| is false, \verb|pkaleu| contains the momenta, including those of the initial-state particles.
The third routine \verb|my_wght| returns the weight associated with the most recently generated phase space point, and the final routine \verb|my_collect| takes the full weight, including the integrand, as input for optimization purposes.

The routines above would suffice to deal with $e^+e^-$ scattering.
In case the user would like to generate events for hadron-hadron scattering, initial-state momenta can be provided to \Kaleu\ for each event with
\begin{verbatim}
  call kinem_inst( kin ,discard ,pkaleu )
\end{verbatim}
This routine should be called just before \verb|kaleu_gnrt|.
The array \verb|pkaleu| has the same format as before, but is input here.
The entries \verb|pkaleu(0:3,-2)| and \verb|pkaleu(0:3,-1)| are taken as the initial-state momenta.
The logical \verb|discard| is output, and is true if something is wrong with the momenta.
Alternatively, the user can let \Kaleu\ deal with the $x$-variables within collinear factorization.
For this option, one item should be added to the module, and a few routine calls should be added.
An example is given in \Appendix{App:strf}.
The optimization of the generation of the $x$-variables is performed with the help of {\sc Parni}.

An example of the use of more instances of the program is given in \Appendix{App:oop}.
%

%%%%%%%%%%%%%%%%%%%%%%%%%%%%%%%%%%%%%%%%%%%%%%%%%%%%%%%%%%%%%%%%%%%%%%%%%%%%%%%%
\section{Results\label{Sec:Perf}}
%%%%%%%%%%%%%%%%%%%%%%%%%%%%%%%%%%%%%%%%%%%%%%%%%%%%%%%%%%%%%%%%%%%%%%%%%%%%%%%%
In order to assess the performance and applicability of \Kaleu, some results existing in literature are reproduced.

\Table{top}, \Table{bosons2}, \Table{higgs2} and \Table{triple1} show some results for $e^{\p}e^{\m}\to6f$ processes as presented in \AmHel.
It concerns all processes at $\sqrt{s}=500\,\GeV$, and ``with QCD'' if applicable.
The numbers in the columns ``{\sc Amegic++}'' and ``\HelPhe'' are taken from that paper.
The numbers in the column ``\HelKal'' were obtained by replacing {\sc Phegas} with \Kaleu\ in the {\sc Helac/Phegas} system.
In \AmHel\, results where obtained using $10^6$ phase space points before cuts.
It is, however, not clear whether this includes an optimization phase.
The results with \Kaleu\ were obtained with $10^7$ phase space points before cuts, {\em including\/} optimization.
In order to compare the performances, both the standard deviation $\sigma$ and $\sqrt{10}\,\sigma$ are presented explicitly for the latter.
Some results with ``\HelPhe'' have been rounded further than in the original paper for clarity.
This leads to an ambiguity for one value, which has been highlighted.
The optimization was performed in $10$ steps of $50\times10^3$ phase space points after cuts.
These were not included in the estimation of the cross section.
The last column in the tables presents the total number of phase space points passing the cuts and leading to a matrix element evaluation, including the $500\times10^3$ evaluations used for optimization.

\begin{table}%[h] 
{\small
\begin{center}
\begin{tabular}{lccccc} 
\hline\noalign{\smallskip}      
Final state          & {\sc Amegic++}  & \HelPhe & \HelKal($\stens$)($\ones$) & $\Nacc$\\
\noalign{\smallskip}\hline\noalign{\smallskip}
 $b\bar b\, u \bar d\,d \bar u$                           %bbudud1 
                     & 49.74(21) & 50.20(13) & 50.33(17)(05) &  7.186e+6 \\
 $b\bar b\, u \bar u\,g g$                                %bbuugg_
                     & 9.11(13)  & 8.83(04) & 8.80(09)(03) &   4.846e+6 \\
 $b\bar b\, g gg g$                                       %bbgggg_
                     & 24.09(18) & 23.80(17) &  23.80(33)(10) &   3.627e+6 \\
 $b\bar b\, u \bar d\, e^{\m} \bar \nu_e$                 %bbuden1
                     & 17.486(66) & 17.492(41) & 17.527(58)(18) &  8.357e+6 \\
 $b\bar b\, e^{\p} \nu_e\, e^{\m} \bar \nu_e$             %bbenen_
                     & 5.954(55) & 5.963(11) & 5.938(18)(06) &  8.453e+6 \\
 $b\bar b\, e^{\p} \nu_e\, \mu^{\m} \bar \nu_{\mu}$       %bbenmn_
                     & 5.865(24) & 5.868(10) & 5.864(19)(06) &  7.812e+6 \\
 $b\bar b\, \mu^{\p} \nu_{\mu}\, \mu^{\m} \bar \nu_{\mu}$ %bbmnmn_
                     & 5.840(30) & 5.839(12) &  5.829(19)(06) &  8.069e+6 \\
\noalign{\smallskip}\hline
\end{tabular}
\end{center}
}
\caption{\label{top}
Cross sections in [fb] for some $e^+\,e^- \to 6f$ processes as in \AmHel. The numbers in the column ``{\sc Amegic++}'' and ``\HelPhe'' are taken directly from that paper. The number $\Nacc$ refers to the run with ``\HelKal''.}
\end{table}
\begin{table}%[h] 
{\small
\begin{center}
\begin{tabular}{l|lccccc}  
\hline\noalign{\smallskip}      
& Final state          & {\sc Amegic++}  & \HelPhe  & \HelKal($\stens$)($\ones$) & $\Nacc$ \\
\noalign{\smallskip}\hline\noalign{\smallskip}
      & $e^{\m}e^{\p}\,u \bar u\,d \bar d$                   %eeuudd11
                     & 1.237(15)&1.265(05) & 1.274(25)(08) &  5.625e+6 \\
      & $e^{\m}e^{\p}\,u \bar u\,e^{\m} e^{\p}$              %eeuuee_1
                     & 6.58(23)e-3 &6.61(08)e-3 &  6.55(41)(13)e-3 &  4.364e+6 \\
with  & $e^{\m}e^{\p}\,u \bar u\,\mu^{\m} \mu^{\p}$          %eeuumm_1
                     & 9.25(17)e-3 &9.1\fbox{5}(07)e-3 & 9.17(18)(06)e-3 &  4.164e+6 \\
Higgs & $\nu_e\bar \nu_e\,u \bar d\,d \bar u$                %nnuddu11
                     & 2.36(7)&2.43(1) &  2.46(3)(1) &  7.141e+6 \\
      & $\nu_e\bar \nu_e\,u \bar d\,e^{\m} \bar \nu_e$       %nnuden_1
                     & 0.916(30) & 0.912(05)  & 0.910(16)(05) &   7.660e+6 \\
      & $\nu_e\bar \nu_e\,u \bar d\,\mu^{\m} \bar \nu_{\mu}$ %nnudmn_1
                     & 0.878(27)&0.889(05) & 0.897(11)(03) &  8.516e+6 \\
\noalign{\smallskip}\hline\noalign{\smallskip}
      & $e^{\m}e^{\p}\,u \bar u\,d \bar d$                   %eeuudd10
                          & 1.0514(97) &1.0445(51) & 1.0561(68)(21) &  5.076e+6 \\
      & $e^{\m}e^{\p}\,u \bar u\,e^{\m} e^{\p}$              %eeuuee_0
                     & 4.082(56)e-3 &4.214(46)e-3 & 4.174(77)(24)e-3 &  3.377e+6 \\
no    & $e^{\m}e^{\p}\,u \bar u\,\mu^{\m} \mu^{\p}$          %eeuumm_0
                     & 5.805(67)e-3 &5.828(49)e-3 & 5.852(83)(26)e-3 &  2.991e+6 \\
Higgs & $\nu_e\bar \nu_e\,u \bar d\,d \bar u$                %nnuddu10
                     & 0.4755(21)  &0.4711(24) &  0.4745(18)(06) &  7.135e+6 \\
      & $\nu_e\bar \nu_e\,u \bar d\,e^{\m} \bar \nu_e$       %nnuden_0
                     & 0.16033(63) &0.16011(78) &  0.16123(47)(15) &  9.018e+6 \\
      & $\nu_e\bar \nu_e\,u \bar d\,\mu^{\m} \bar \nu_{\mu}$ %nnudmn_0
                     & 0.14383(53) &0.14439(65) &  0.14367(36)(11) &  9.239e+6 \\
\noalign{\smallskip}\hline
\end{tabular}
\end{center}
}
\caption{\label{bosons2}
Cross sections in [fb] for some $e^+\,e^- \to 6f$ processes as in \AmHel. The numbers in the column ``{\sc Amegic++}'' and ``\HelPhe'' are taken directly from that paper. The number $\Nacc$ refers to the run with ``\HelKal''.}
\end{table}
\begin{table}%[h] 
{\small
\begin{center}
\begin{tabular}{l|lccccc} 
\hline\noalign{\smallskip}              
 & Final state          & {\sc Amegic++}  &\HelPhe  &  \HelKal($\stens$)($\ones$) & $\Nacc$\\
\noalign{\smallskip}\hline\noalign{\smallskip}
      & $\mu^{\m}\mu^{\p}\,\mu^{\m} \bar \nu_{\mu}\,e^{\p}\nu_e$ %mmmnen_1 
                     & 0.03747(29) &0.03749(32) &  0.03740(31)(10) &  4.868e+6 \\
with  & $\mu^{\m}\mu^{\p}\,u \bar d\,e^{\m} \bar \nu_e$                %mmuden_1
                     & 0.1106(22) &0.1090(07) & 0.1083(10)(03) &  5.245e+6 \\
Higgs & $\mu^{\m}\mu^{\p}\,\mu^{\m}  \mu^{\p}\,e^{\m} e^{\p}$          %mmmmee_1
                     & 2.731(065)e-3 &2.691(042)e-3 &  2.737(108)(034)e-3 &  2.285e+6 \\
      & $\mu^{\m}\mu^{\p}\,u \bar u\,d \bar d$                         %mmuudd11
                     & 0.2634(22) &0.2642(15) & 0.2638(22)(07) &  5.218e+6 \\
      & $\mu^{\m}\mu^{\p}\,u \bar u\,u \bar u$                         %mmuuuu11
                     & 8.767(65)e-3  &8.978(58)e-3 &  8.882(87)(28)e-3 &   3.614e+6 \\
\noalign{\smallskip}\hline\noalign{\smallskip}
      & $\mu^{\m}\mu^{\p}\,\mu^{\m} \bar \nu_{\mu}\,e^{\p}\nu_e$ %mmmnen_0
                     & 0.03054(23) &0.03092(19) & 0.03079(22)(07) &  4.512e+6 \\
no    & $\mu^{\m}\mu^{\p}\,u \bar d\,e^{\m} \bar \nu_e$                %mmuden_0
                     & 0.08911(53) &0.08925(48) &  0.08932(51)(16) &  4.800e+6 \\
Higgs & $\mu^{\m}\mu^{\p}\,\mu^{\m}  \mu^{\p}\,e^{\m} e^{\p}$          %mmmmee_0
                     & 2.280(66)e-3 &2.277(62)e-3 &  2.224(82)(26)e-3 &   1.847e+6 \\
      & $\mu^{\m}\mu^{\p}\,u \bar u\,d \bar d$                         %mmuudd10
                     & 0.2092(12) &0.2075(13) & 0.2092(10)(03) &  5.152e+6 \\
      & $\mu^{\m}\mu^{\p}\,u \bar u\,u \bar u$                         %mmuuuu10
                     & 6.134(29)e-3 &6.108(27)e-3 & 6.072(41)(13)e-3 &  3.907e+6 \\
\noalign{\smallskip}\hline
\end{tabular}
\end{center}
}
\caption{\label{higgs2}
Cross sections in [fb] for some $e^+\,e^- \to 6f$ processes as in \AmHel. The numbers in the column ``{\sc Amegic++}'' and ``\HelPhe'' are taken directly from that paper. The number $\Nacc$ refers to the run with ``\HelKal''.}
\end{table}
\begin{table}%[h] 
{\small
\begin{center}
\begin{tabular}{l|lccccc} 
\hline\noalign{\smallskip}              
 & Final state          & {\sc Amegic++}  & \HelPhe  & \HelKal($\stens$)($\ones$) & $\Nacc$ \\
\noalign{\smallskip}\hline\noalign{\smallskip}
with Higgs & $\mu^{\m}\mu^{\p}\,b \bar bb \bar b$ %mmbbbb_1
                     & 30.96(0.60)e-3 &30.19(0.43)e-3 & 29.96(1.42)(0.45)e-3 &   3.153e+6 \\
\noalign{\smallskip}\hline\noalign{\smallskip}
no Higgs & $\mu^{\m}\mu^{\p}\,b \bar bb \bar b$ %mmbbbb_0
                     & 6.308(24)e-3 & 6.364(21)e-3 & 6.377(26)(08)e-3 &  4.918e+6 \\
\noalign{\smallskip}\hline
\end{tabular}
\end{center}
}
\vspace{-10pt}
\caption{\label{triple1}
Cross sections in [fb] for some $e^+\,e^- \to 6f$ processes as in \AmHel. The numbers in the column ``{\sc Amegic++}'' and ``\HelPhe'' are taken directly from that paper. The number $\Nacc$ refers to the run with ``\HelKal''.}
\end{table}

\Table{8fermion} shows some results for $e^{\p}e^{\m}\to8f$ processes as presented in \Carlo.
It concerns some processes with $m_{H}=130\,\GeV$ and $m_t=174.3\,\GeV$, at $\sqrt{s}= 800\,\GeV$ and with QCD.
%: leptonic, semileptonic and hadronic, for each category the process from \Carlo\ with the least and the most number of Feynman graphs.
%
The numbers in the column ``{\sc Carlomat}'' are directly taken from that paper.
The numbers in the column ``{\sc Helac/Kaleu}'' were again obtained by replacing {\sc Phegas} with \Kaleu\ in the {\sc Helac/Phegas} system.
They were obtained with $10^8$ phase space points before cuts, including optimization.
The optimization was performed in $10$ steps of $100\times10^3$ phase space points after cuts, which were not included in the estimation of the cross section.
The last column in the tables presents the total number of phase space points passing the cuts and leading to a matrix element evaluation. 

\begin{table}%[h] 
{\small
\begin{center}
\vspace{10pt}
\begin{tabular}{lcccc} 
\hline\noalign{\smallskip}      
Final state          & {\sc Carlomat} & {\sc Helac/Kaleu} & $\Nacc$\\
\noalign{\smallskip}\hline\noalign{\smallskip}
 $b\bar b\,b\bar b\, \nu_{\mu}\mu^{\p}\,\tau^{\m}{\bar\nu}_\tau$ %bbbbmntn11
                     & 35.9(1) & 36.17(15) & 18.12e+6\\
 $b\bar b\,b\bar b\, \nu_e e^{\p}\,e^{\m}{\bar\nu}_e$ %bbbbenen11
                     & 36.1(1) & 36.37(18) & 16.82e+6 \\
 $b\bar b\,b\bar b\, u\bar d\,\mu^{\m}{\bar\nu}_\mu$ %bbbbudmn11
                     & 100.6(2) & 100.50(46) & 13.21e+6 \\
 $b\bar b\,b\bar b\, c\bar s\,e^{\m}{\bar\nu}_e$ %bbbbcsen11
                     & 100.5(3) & 100.52(47) & 12.63e+6 \\
 $b\bar b\,b\bar b\, u\bar d\,s\bar c$ %bbbbudsc11
                     & 314(2) & 314.1(1.4) & 13.50e+6 \\
 $b\bar b\,b\bar b\, u\bar d\,d\bar u$ %bbbbcssc11
                     & 314(1) & 320.1(1.9) & 13.36e+6 \\
\noalign{\smallskip}\hline
\end{tabular}
\end{center}
}
\vspace{-10pt}
\caption{\label{8fermion}
Cross sections in [ab] for some $e^+\,e^- \to 8f$ processes as in \Carlo. The numbers in the column ``{\sc Carlomat}'' are taken directly from that paper. The number $\Nacc$ refers to the run with ``{\sc Helac/Kaleu}''.}
\end{table}

In order to present some results for hadron-hadron collisions, \Kaleu\ has been connected with \Alpgen~\Alpcite.
This program deals with partonic subprocesses in hadron-hadron collisions in one Monte Carlo run, making an optimized choice of a subprocess to be generated for each event.
Because \Kaleu\ can provide an independent generator for each subprocess, it can easily be merged into the structure of \Alpgen.
This has been done such, that \Alpgen\ still makes the choice of a subprocess for each event, and that, given the subprocess, \Kaleu\ generates the variables $x_1,x_2$ for the PDFs and the final-state momenta.
It should be noted, however, that for the purpose of this write-up, the connection between the two programs has been established only to the level of cross section calculation and not, for example, to full event un-weighting.
\Table{ppcoll} presents results for processes of the type $pp\to{}t\bar t+N\mathrm{jets}$ and $pp\to{}t\bar t\,b\bar b+N\mathrm{jets}$.
The numbers in the column ``{\sc Alpgen}'' are taken directly from \Alpcite.
The user of \Alpgen\ chooses the number of matrix element evaluations for a Monte Carlo run, not the number of generated phase space points before cuts.
The last column presents the number of evaluations for the run with \Kaleu, including those used for optimization but omitted for the estimation of the cross section.

\begin{table}%[h] 
{\small
\begin{center}
\begin{tabular}{lcccc} 
\hline\noalign{\smallskip}      
Final state          & {\sc Alpgen} & {\sc Alpgen/Kaleu} & $\Nacc$\\
\noalign{\smallskip}\hline\noalign{\smallskip}
 $t\bar t+2\mathrm{jets}$ & 255(1)   & 254.38(73) & 11.4e+6 \\
 $t\bar t+3\mathrm{jets}$ & 111.5(5) & 111.09(43) & 11.4e+6 \\
 $t\bar t+4\mathrm{jets}$ & 42.4(4)  & 42.72(45) & 11.6e+6 \\
 $t\bar t+5\mathrm{jets}$ & 14.07(16)& 14.36(13) & 11.6e+6 \\
 $t\bar t+6\mathrm{jets}$ & 4.36(8)  & 4.369(43) & 13.0e+6 \\
 $t\bar t\,b\bar b$                & 1.35(1) & 1.3490(21) & 10.2e+6 \\
 $t\bar t\,b\bar b+1\mathrm{jet}$  & 1.47(2) & 1.4624(42) & 10.4e+6 \\
 $t\bar t\,b\bar b+2\mathrm{jets}$ & 0.94(2)  & 0.9280(40) & 10.6e+6 \\
 $t\bar t\,b\bar b+3\mathrm{jets}$ & 0.457(8) & 0.4522(28) & 10.6e+6 \\
 $t\bar t\,b\bar b+4\mathrm{jets}$ & 0.189(4) &  0.1851(14)& 6.2e+6 \\
\noalign{\smallskip}\hline
\end{tabular}
\end{center}
}
\caption{\label{ppcoll}
Cross sections in [pb] for some $pp$ collision processes at LHC as in \Alpcite. The numbers in the column ``{\sc Alpgen}'' are taken directly from that paper. The number $\Nacc$ refers to the run with ``{\sc Alpgen/Kaleu}''.}
\end{table}

%%%%%%%%%%%%%%%%%%%%%%%%%%%%%%%%%%%%%%%%%%%%%%%%%%%%%%%%%%%%%%%%%%%%%%%%%%%%%%%%
\section{Summary}
%%%%%%%%%%%%%%%%%%%%%%%%%%%%%%%%%%%%%%%%%%%%%%%%%%%%%%%%%%%%%%%%%%%%%%%%%%%%%%%%
The program \Kaleu\ for parton-level phase space generation was presented.
Given the masses and widths of all real and virtual particles, it generates importance sampled phase space points for the scattering process under consideration.
Given the total weight of each event, including the value of the integrand, it optimizes further to this integrand on the fly.
It is written in {\sc Fortran}, such that it can independently deal with several scattering processes in parallel.

%%%%%%%%%%%%%%%%%%%%%%%%%%%%%%%%%%%%%%%%%%%%%%%%%%%%%%%%%%%%%%%%%%%%%%%%%%%%%%%
\subsection*{Acknowledgments}
%%%%%%%%%%%%%%%%%%%%%%%%%%%%%%%%%%%%%%%%%%%%%%%%%%%%%%%%%%%%%%%%%%%%%%%%%%%%%%%
The author would like to thank C.G.~Papadopoulos for useful discussions.

%%%%%%%%%%%%%%%%%%%%%%%%%%%%%%%%%%%%%%%%%%%%%%%%%%%%%%%%%%%%%%%%%%%%%%%%%%%%%%%

%%%%%%%%%%%%%%%%%%%%%%%%%%%%%%%%%%%%%%%%%%%%%%%%%%%%%%%%%%%%%%%%%%%%%%%%%%%%%%%

%\newpage
%%%%%%%%%%%%%%%%%%%%%%%%%%%%%%%%%%%%%%%%%%%%%%%%%%%%%%%%%%%%%%%%%%%%%%%%%%%%%%%
\begin{appendix}
%%%%%%%%%%%%%%%%%%%%%%%%%%%%%%%%%%%%%%%%%%%%%%%%%%%%%%%%%%%%%%%%%%%%%%%%%%%%%%%
\section{Generation of $x$-variables\label{App:strf}}
%%%%%%%%%%%%%%%%%%%%%%%%%%%%%%%%%%%%%%%%%%%%%%%%%%%%%%%%%%%%%%%%%%%%%%%%%%%%%%%
An example of user-defined routines which can be used for the generation of events for hadron-hadron scattering within collinear factorization.

{\scriptsize
\begin{verbatim}
module my_kaleu
  use avh_kaleu
  use avh_kaleu_model
  use avh_kaleu_kinem
  type(model_type) ,save :: mdl ! Masses and widths
  type(kinem_type) ,save :: kin ! Kinematics
  type(kaleu_type) ,save :: obj ! Instance of the phase space generator
  type(strf_type)  ,save :: str ! Concerns the generation of x1,x2
end module
      
subroutine my_init( process ,nfinst ,ecm ,nbatch,nstep,thrs )
  use my_kaleu
  integer ,intent(in)         :: process(-2:17),nfinst,nbatch,nstep
  real(kind(1d0)) ,intent(in) :: ecm,thrs
  type(vertx_type) :: vtx
  call my_model( mdl,vtx )
  call kaleu_put_process( mdl,vtx,kin,obj ,process ,nfinst ,ecm )
  call kaleu_init_adapt( mdl,obj ,nbatch,nstep,thrs )
  call kaleu_init_strf( str,kin ,0d0 )
end subroutine
  
subroutine my_gnrt( discard ,x1kaleu,x2kaleu,pkaleu )
  use my_kaleu
  logical         ,intent(out) :: discard
  real(kind(1d0)) ,intent(out) :: ,x1kaleu,x2kaleu,pkaleu(0:3,-2:17)
  call kaleu_gnrt_strf( str,kin ,discard ,x1kaleu,x2kaleu )  
  call kaleu_gnrt( mdl,kin,obj ,discard ,pkaleu )
end subroutine

subroutine my_wght( weight )
  use my_kaleu
  real(kind(1d0)) ,intent(out) :: weight
  real(kind(1d0)) :: ww
  call kaleu_wght( mdl,kin,obj ,weight )
  call kaleu_wght_strf( str ,ww )
  weight = weight*ww
end subroutine

subroutine my_collect( weight )
  use my_kaleu
  real(kind(1d0)) ,intent(in) :: weight
  call kaleu_collect( obj ,weight )
  call kaleu_collect_strf( str ,weight )
end subroutine
\end{verbatim}
}
\noindent The last argument of \verb|kaleu_init_strf| is a minimum value for an $x$-variable.
If this number is smaller than the number dictated by kinematical limits, the latter is used.
The output \verb|x1kaleu| and \verb|x2kaleu| of the generation routine are the values at which the PDFs should be evaluated.

%%%%%%%%%%%%%%%%%%%%%%%%%%%%%%%%%%%%%%%%%%%%%%%%%%%%%%%%%%%%%%%%%%%%%%%%%%%%%%%
\section{Example of the use of more instances\label{App:oop}}
%%%%%%%%%%%%%%%%%%%%%%%%%%%%%%%%%%%%%%%%%%%%%%%%%%%%%%%%%%%%%%%%%%%%%%%%%%%%%%%
As an example of the use of more instances of the program, the user could want to treat several subprocesses in one Monte Carlo run, choosing another subprocess for each event.
Then the user-defined routines could get one more integer input variable \verb|iproc| determining the subprocess, and could look like

{\scriptsize
\begin{verbatim}
module my_kaleu
  use avh_kaleu
  use avh_kaleu_model
  use avh_kaleu_kinem
  integer ,parameter :: nmax = 20;
  type(model_type) ,save :: mdl       ! Masses and widths
  type(kinem_type) ,save :: kin(nmax) ! Kinematics
  type(kaleu_type) ,save :: obj(nmax) ! Instance of the phase space generator
  type(strf_type)  ,save :: str(nmax) ! Concerns the generation of x1,x2
end module
      
subroutine my_init( iproc ,process ,nfinst ,ecm ,nbatch,nstep,thrs )
  use my_kaleu
  integer ,intent(in)         :: iproc,process(-2:17),nfinst,nbatch,nstep
  real(kind(1d0)) ,intent(in) :: ecm,thrs
  type(vertx_type) :: vtx
  call my_model( mdl,vtx )
  call kaleu_put_process( mdl,vtx,kin(iproc),obj(iproc) ,process ,nfinst ,ecm )
  call kaleu_init_adapt( mdl,obj(iproc) ,nbatch,nstep,thrs )
  call kaleu_init_strf( str(iproc),kin(iproc) ,0d0 )
end subroutine
  
subroutine my_gnrt( iproc ,discard ,x1kaleu,x2kaleu,pkaleu )
  use my_kaleu
  integer         ,intent(in)  :: iproc
  logical         ,intent(out) :: discard
  real(kind(1d0)) ,intent(out) :: ,x1kaleu,x2kaleu,pkaleu(0:3,-2:17)
  call kaleu_gnrt_strf( str(iproc),kin(iproc) ,discard ,x1kaleu,x2kaleu )  
  call kaleu_gnrt( mdl,kin(iproc),obj(iproc) ,discard ,pkaleu )
end subroutine

subroutine my_wght( iproc ,weight )
  use my_kaleu
  integer         ,intent(in)  :: iproc
  real(kind(1d0)) ,intent(out) :: weight
  real(kind(1d0)) :: ww
  call kaleu_wght( mdl,kin(iproc),obj(iproc) ,weight )
  call kaleu_wght_strf( str(iproc) ,ww )
  weight = weight*ww
end subroutine

subroutine my_collect( iproc ,weight )
  use my_kaleu
  integer         ,intent(in) :: iproc
  real(kind(1d0)) ,intent(in) :: weight
  call kaleu_collect( obj(iproc) ,weight )
  call kaleu_collect_strf( str(iproc) ,weight )
end subroutine
\end{verbatim}
}

\end{appendix}
%%%%%%%%%%%%%%%%%%%%%%%%%%%%%%%%%%%%%%%%%%%%%%%%%%%%%%%%%%%%%%%%%%%%%%%%%%%%%%%
%
%
\end{document}